\documentclass[amsmath,amssymb,amsfonts,aps,pre,superscriptaddress,bibnotes,showpacs,showkeys,longbibliography,10pt,twocolumn]{revtex4-1}

\usepackage{epsfig}
\usepackage{ulem}
\usepackage[english]{babel}
\usepackage{color}
\usepackage{hyperref}

\begin{document}
\title{Wetting and nonwetting near a tricritical point}

\author{Joseph O. Indekeu}
\affiliation{Institute for  Theoretical Physics, KU Leuven, BE-3001 Leuven, Belgium}
\author{Kenichiro Koga}
\affiliation{Research Institute for Interdisciplinary Science, Okayama University, Okayama 700-8530, Japan}

\date{\today}

\begin{abstract}
The dihedral contact angles between interfaces in three-fluid-phase equilibria must be continuous functions of the bulk thermodynamic fields. This general argument, which we propose, predicts a nonwetting gap in the phase diagram, challenging the common belief in ``critical-point wetting", even for short-range forces. A demonstration is provided by exact solution of a mean-field two-density functional theory for three-phase equilibria near a tricritical point (TCP). Complete wetting is found in a tiny vicinity of the TCP. Away from it, nonwetting prevails and no wetting transition takes place, not even when a critical endpoint is approached. Far from the TCP, reentrant wetting may occur, with a different wetting phase. These findings shed light on hitherto unexplained experiments on ternary ${\rm H_2O}$-oil-nonionic amphiphile mixtures in which nonwetting continues to exist as one approaches either one of the two critical endpoints.
\end{abstract}

\maketitle
Wetting phenomena and in particular wetting phase transitions have enjoyed much attention from physicists and other scientists or engineers.
The first theoretical predictions and experimental observations of a phase transition from a state of partial wetting (``nonwet''), in which three phases are pairwise in contact at their mutual interfaces and meet at a common contact line, to a state of complete wetting (``wet''), in which one phase intrudes at the interface between the other two, date from the late 1970s \cite{Cahn,ES,MC}. Not only first-order but also critical wetting transitions (of second or higher order) may occur \cite {Sul,NF}. Numerous reviews have since been dedicated to the development of this fascinating field, e.g., \cite{dG,SulTdG,Diet,FLN,BonnRoss,BonnRMP}.

In 1977 Cahn predicted that complete wetting must be expected upon approach of a critical point where two phases become identical. This is termed ``critical-point wetting'' (CPW) \cite{Cahn}. Subsequently, insight was gained in which systems CPW does occur and in which it does not. For short-range forces, CPW must take place when two coexisting phases adsorbed at a wall with a nonzero ``surface field'' approach bulk criticality, as found in \cite{NF} and \cite{PNAS} (surface phase transition class C defined in \cite{PNAS}). For a vanishing surface field CPW is suppressed \cite{Ind,Sevrin,Igloi}. For long-range wall-fluid and/or fluid-fluid forces, nonwetting gaps exist in which CPW does not occur, as predicted in \cite{NighInd,EbnerSaam} and recently demonstrated in \cite{PNAS} (classes A, B and D defined in \cite{PNAS}). 

In this Letter we ask whether CPW should be expected when three coexisting fluid phases are treated on equal footing instead of replacing one of them by an {\it ad hoc} ``wall" boundary condition. We consider molecular fluids governed by van der Waals forces, or fluids with forces of shorter range, such as plasmas and electrolytes \cite{dG} or colloid-polymer mixtures \cite{Lek}. CPW can be tested approaching a critical endpoint (CEP) where two (out of three) phases become identical and the resulting critical phase coexists with the third, noncritical, phase. Especially relevant is the vicinity of a tricritical point (TCP) at which two CEP lines meet, in accord with Gibbs' phase rule. Pioneering experiments in this arena were carried out by Widom {\it et al.} \cite{LW,NWW} and \cite{NWW} was judged to be consistent with CPW \cite{ZinnFisher}. However, in ternary H$_2$O-oil-nonionic amphiphile mixtures nonwetting was observed to persist  \cite{KahlweitBusse}.

To shed light on these issues, we consider a mixture with three components and study three-phase equilibria with coexisting phases $\alpha$, $\beta$ and $\gamma$ in mean-field density-functional theory (DFT) with two spatially varying densities $\rho_1$ and $\rho_2$. Two densities are necessary because near a TCP, in a single-density theory the ``middle'' phase always wets the interface between the other two phases \cite{Widom,RW}. Gibbs' phase rule dictates that two linear combinations of pressure $p$, temperature $T$ and chemical potentials $\mu_i$, $i=1,2,3$, can be taken as independent field coordinates, say $s$ and $t$, in the two-dimensional three-phase coexistence surface in thermodynamic space. This surface is bounded by two CEP lines that meet tangentially (with a power $3/2$) at the TCP, located at $s=t=0$, along an asymptote that we take to be the $t$-axis. We can use $t\geq 0$ to measure the ``temperature distance" to the TCP , in the $(s,t)$-plane, and use $s$ to interpolate between two CEPs at fixed $t$ \cite{Grif,LW,RW}. 

Of chief interest here are fluids with short-range forces, for which the single-density approximation with a wall boundary condition always predicts CPW. The model we employ is a square-gradient DFT akin to model ${\rm T}$ in \cite{KW}, in which the bulk free-energy density is a product of three potential wells, centered about different points in the $(\rho_1,\rho_2)$-plane, with bulk phase density pairs $(\rho_{1}^{\nu},\rho_{2}^{\nu})$, with $ \nu = \alpha, \beta,\gamma$. In the theory of tricritical phenomena due to Griffiths an important constraint reflects the asymptotic scaling properties of the three-phase region close to the TCP \cite{Grif,RW}. Using linear combinations of the densities, which we rename to be just our $\rho_1$ and $\rho_2$, the constraint on the bulk densities, in each phase, takes the form
\begin{equation}\label{constraint}
\rho^{\nu}_2= -(\rho^{\nu}_1)^2.
\end{equation} 
The (dimensionless) interfacial free-energy density $\Psi$  is taken to be
\begin{equation}\label{Psi}
\Psi (\rho_1,\rho_2) = \frac{1}{2} \sum_{i=1,2}\left (\frac{d\rho_i}{dz}\right )^2   + \prod_{\nu=\alpha,\beta,\gamma}\, \sum_{i=1,2}(\rho_i - \rho_{i}^{\nu} )^2. \end{equation}
This simple symmetric form has been chosen in the interest of obtaining an exactly solvable DFT. Additional parameters may affect the (order and locus of the) wetting transitions, as they do in \cite{KIW}, but the presence of a nonwetting gap, for which we will give a general argument, ought to be robust. 

The bulk free-energy density is minimal and takes equal values (chosen to be zero) in all three bulk phases. The densities $\rho_1(z)$ and $\rho_2(z)$, with $z$ the coordinate perpendicular to the interface between any two phases, characterize the structure of that interface. The equilibrium (or ``optimal'') densities minimize the excess free-energy functional associated with the interface,
\begin{equation} \label{sigma}
\sigma [\rho_1,\rho_2] = \int_{-\infty}^{\infty}dz \;\Psi (\rho_1,\rho_2),
\end{equation}
subject to the boundary conditions that the interface connects two (spatially homogeneous) bulk phases that are a macroscopic distance apart. Therefore we impose one of the bulk phases, e.g.,  $\alpha$, at $z= -\infty$  and the other, e.g., $\beta$, at $z=\infty$. The equilibrium interfacial tension $\sigma_{\alpha\beta}$ is then the value of $\sigma [\rho_1,\rho_2]$ for the optimal density profiles. 

The densities $\rho_1$ for the three coexisting phases, calculated to be the zeroes of the third-degree polynomial
$P (\rho_1) = \rho_1^3-3t\rho_1+2s$, are ordered in the manner $\rho_1^{\alpha}(s,t) \leq \rho_1^{\beta}(s,t) \leq \rho_1^{\gamma}(s,t)$ \cite{RW}. 
At fixed $t$, $s$ can interpolate between two CEPs, one at $\alpha\beta$ criticality ($ s\,t^{-3/2}=-1$) and one at $\beta\gamma$ criticality ($ s\,t^{-3/2}=1$). The results for $s<0$ can be obtained from those for $s>0$ by interchanging phases $\alpha$ and $\gamma$, so we consider $s \geq 0$ and investigate the range $0\leq st^{-3/2} \leq 1$. 

There are now two possibilities. Either the $\beta$ phase does not wet the $\alpha\gamma$ interface, in which case
\begin{equation}\label{nonwet}
\sigma_{\alpha\gamma} < \sigma_{\alpha\beta}  + \sigma_{\beta\gamma} , \;\; \mbox{nonwet},
\end{equation}
or the $\beta$ phase wets the $\alpha\gamma$ interface, and then,
\begin{equation}\label{wet}
\sigma_{\alpha\gamma} = \sigma_{\alpha\beta}  + \sigma_{\beta\gamma} , \;\; \mbox{wet},
\end{equation}
the latter of which is sometimes referred to as ``Antonov's rule" \cite{RW}. When $\beta$ does not wet the $\alpha\gamma$ interface, it is possible that $\gamma$ wets the $\alpha\beta$ interface (or that it does not). 

We take a twofold approach. High-precision numerical integration of the free-energy density is performed to obtain the interfacial tensions. Also, recently conjectured simple analytic forms of these interfacial tensions \cite{KogaIndekeu} are used to calculate them exactly. The two methods provide indistinguishable results, while the analytic calculation is by far the simplest. 

The analytic forms allow one to obtain a geometrical representation of the wetting criterion. To elucidate this we recall the analytic expression for, say, $\sigma_{\alpha\gamma}$ uncovered in \cite{KogaIndekeu}. It is applicable to a three-phase triangle of arbitrary geometry and valid for a non-wet interface,
\begin{equation}\label{pcubeEll}
\sigma_{\alpha\gamma} = \frac{\sqrt{2}}{6} p^3 \ell,
\end{equation}
with $p$ the Euclidean distance from $\alpha$ to $\gamma$ and $\ell$ the Euclidean distance from $\beta$ to the midpoint of the $\alpha\gamma$ line, in the $(\rho_1,\rho_2)$-plane. Note that $p$ is the length of an edge, and $\ell$ that of the conjugate median, in the three-phase triangle.

\begin{figure}[h!]
\centering
\includegraphics[width=0.9\linewidth]{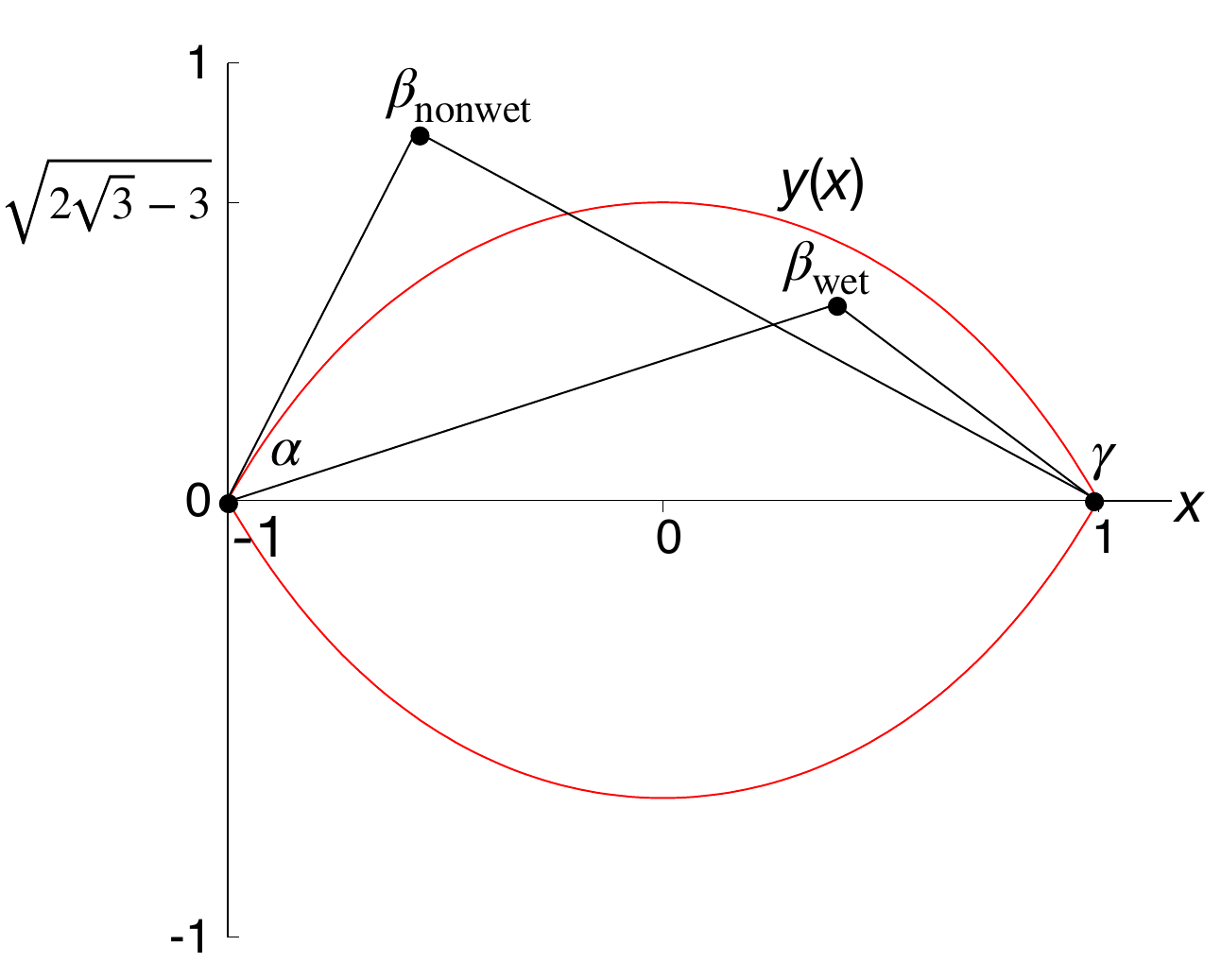}
\caption{Two representative three-phase triangles, after a coordinate transformation from $(\rho_1,\rho_2)$ to $(x,y)$, together with the two curves $y(x)$ that correspond to the exact phase boundaries for wetting of the $\alpha\gamma$ interface by the $\beta$-phase. The curves feature $y(0)^2 = 2 \sqrt{3} -3$ and $|dy/dx|_{x= \pm 1} = \sqrt{3}$. A (non)wet state results if $\beta$ is (outside) inside the domain bounded by the curves.}
\label{1}
\end{figure}

We now construct the geometrical wetting criterion (see Fig.1). We translate, rotate and uniformly rescale the three-phase triangle without affecting its shape, in the $(\rho_1,\rho_2)$-plane.
After this, the $\alpha$-phase point is fixed at $(x,y)=(-1,0)$ and the $\gamma$-phase point at $(x,y)=(1,0)$. Next we express the wetting condition \eqref{wet} in $x$ and $y$, using \eqref{pcubeEll} for each interface, assuming a nonwet state (none of the interfaces between any two phases is wet by the third phase). We find that supposing Antonov's rule is equivalent to drawing two curves $y(x)$, which trace the boundary of the domain of wet states. An involved algebraic calculation, invoking Apollonius' geometrical theorem, shows that the curves satisfy 
\begin{equation}\label{AntonovSimple}
y^2= x^2 -3 +2 \sqrt{x^4 -3x^2 + 3}.
\end{equation}
A state in which the $\alpha\gamma$ interface is wet by $\beta$ results if and only if the $\beta$-phase point lies inside the domain bounded by these curves. The boundary corresponds to the wetting transition. A similar construction applies for wetting by $\alpha$ or $\gamma$. Consequently, a {\it necessary} condition for wetting of an interface between two phases by a third phase is that the former two span the longest edge of the three-phase triangle.

The global wetting phase diagram is displayed in Fig.2. Very close to the TCP, we find what can be called ``tricritical-point wetting"; e.g., for $s=0$ and $t < 2/\sqrt{3}-1 = 0.1547...$, $\beta$ wets the $\alpha\gamma$ interface, in accord with findings in \cite{KW}. The profiles $\rho_1 (z)$ and $\rho_2(z)$, that connect $\alpha$ to $\gamma$ in the $(\rho_1,\rho_2)$-plane, pass through $\beta$. There is no direct $\alpha\gamma$ contact, but a composite $\alpha\gamma$ configuration is found that consists of an $\alpha\beta$ interface, an intruding bulk $\beta$ phase, and a $\beta\gamma$ interface.  

\begin{figure}[h!]
\centering
\includegraphics[width=0.9\linewidth]{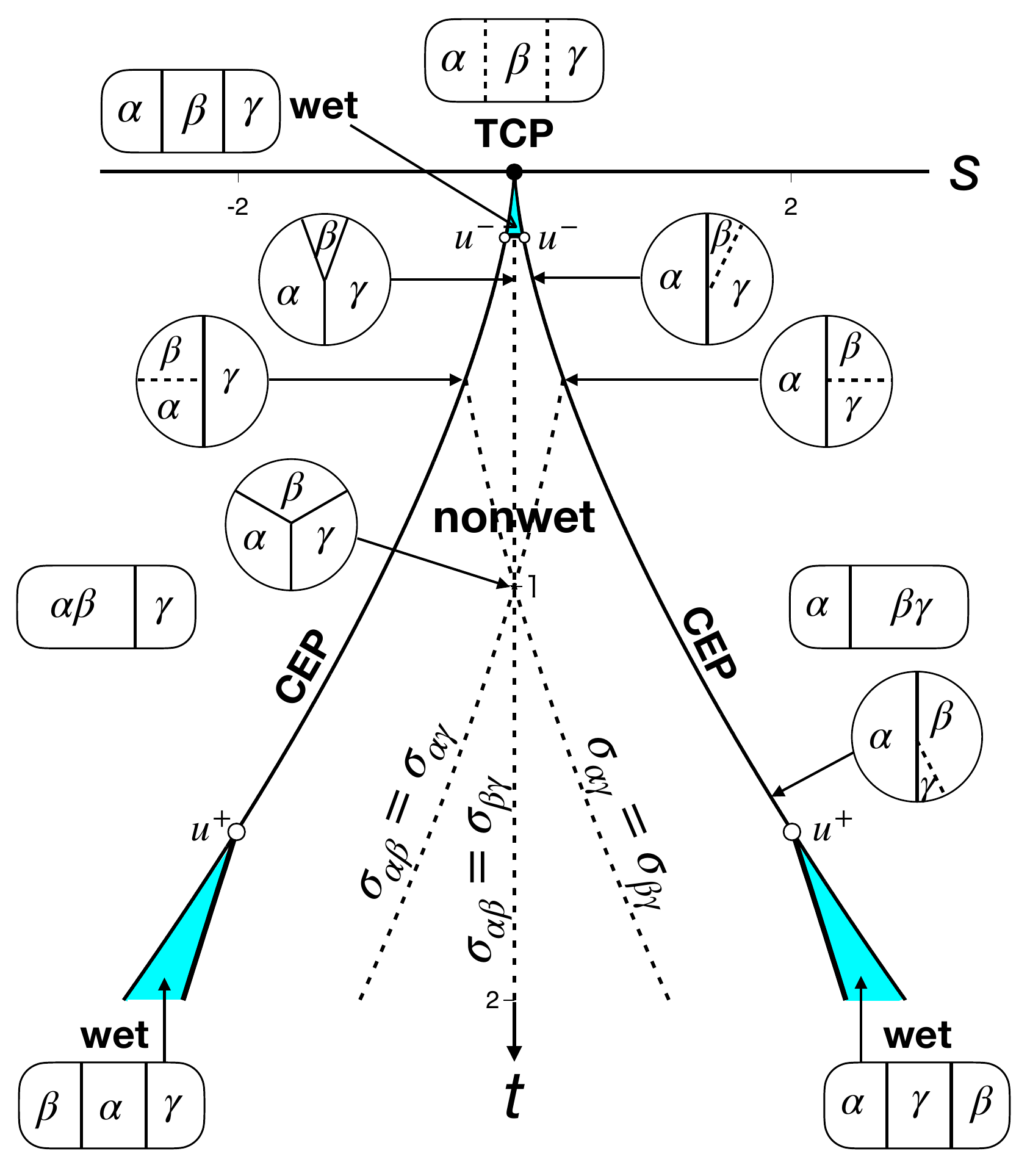}
\caption{Global wetting phase diagram near a TCP. Projected onto the $(s,t)$-plane, the three-phase coexistence region is the surface bounded by the two CEP lines, one for $\alpha\beta$ criticality and one for $\beta\gamma$ criticality. Very close to the TCP a second-order wetting phase boundary is found above which $\beta$ wets the $\alpha\gamma$ interface. Far from the TCP and close to the CEP lines reentrant second-order wetting phase boundaries appear, one for $s<0$, leading to wetting of the $\beta\gamma$ interface by $\alpha$, and one for $s>0$, leading to wetting of the $\alpha\beta$ interface by $\gamma$. The wetting phase boundaries terminate at four points denoted $u^-, u^+$ on the CEP lines, leaving a large gap in $t$, $t^-_u < t < t^+_u$, in which only nonwetting occurs. Also shown are the 3 (dotted) lines of symmetry on each of which two interfacial tensions are equal. The circular cartoons depict three-phase configurations: a nonwet state for $s=0$ close to the TCP, the fully symmetric state at $(s=0,t=1)$, and three nonwetting states infinitesimally close to the CEP line for $\beta\gamma$ criticality. In the latter three the $\beta\gamma$ interface, with vanishing tension, is diffuse (dotted line) and the asymptotic value of $\cos \hat \beta$ along the CEP line is displayed in Fig.3.}
\label{2}
\end{figure}

Very close to the TCP a wetting transition phase boundary is found, with midpoint $(0, 2/\sqrt{3}-1)$ and terminating on the CEP lines at ``unbinding transition" points $u^-$ (discussed later). For $2/\sqrt{3}-1 < t < t_u^- \equiv (7-\sqrt{33})/8 = 0.1569...$ we obtain the locus $t\equiv t_{\rm w}^{\beta}(s)$ of transitions from nonwet to wet states.  Their existence was anticipated in \cite{KW}.

States above this line are wet states near the TCP. States below it feature pairwise direct interfacial contact between $\alpha$ and $\beta$, $\alpha$ and $\gamma$, and $\beta$ and $\gamma$. The three phases meet along a contact line, such that, in a projection onto a plane perpendicular to the line, the three interfaces display dihedral contact angles named after the phase they subtend \cite{RW}, e.g., $\hat\nu$ is the angle for phase $\nu$.

The dihedral angles are related to the interfacial tensions geometrically through Neumann's triangle \cite{KW}. The relevant angle for describing (non-)wetting by $\beta$ is $\hat\beta$, and it satisfies
\begin{equation}\label{cosbeta}
\cos\hat\beta = \frac{1}{2}\left ( \frac{\sigma_{\alpha\gamma}}{\sigma_{\alpha\beta}} \frac{\sigma_{\alpha\gamma}}{\sigma_{\beta\gamma}} - \frac{\sigma_{\alpha\beta}}{\sigma_{\beta\gamma}} - \frac{\sigma_{\beta\gamma}}{\sigma_{\alpha\gamma}}\right ).
\end{equation}

We find that $\cos\hat\beta$ approaches unity proportionally to the second power of the field distance to wetting,
\begin{equation}
\begin{cases}
\cos\hat\beta -1 \propto (t-t_{\rm w}) ^2, \;\; \mbox{for fixed} \; s\\
\cos\hat\beta -1 \propto (s-s_{\rm w}) ^2, \;\; \mbox{for fixed} \; t,
\end{cases}
\end{equation}
with $(s_{\rm w},t_{\rm w})$ a point on $t_{\rm w}^{\beta}(s)$. This is characteristic of a critical wetting transition of second order.

The physics changes drastically when we move away from the TCP, in the broad temperature range $ t_u^- < t < t_u^+\equiv (7 + \sqrt{33})/8 = 1.5930...$. All states are nonwet. Even approaching a CEP, no wetting takes place. For example, for $t_u^- < t < 1/2$ and in the limit $s\,t^{-3/2} \rightarrow 1$, both $\sigma_{\alpha\gamma}-\sigma_{\alpha\beta}$ and $\sigma_{\beta\gamma}$ vanish but their ratio, $\cos\hat\beta$, remains finite and strictly below unity. So the $\alpha\gamma$ interface is not wet by $\beta$. The asymptotic value of $\cos\hat\beta$ varies along the CEP line (see Fig.3). This persistence of nonwetting, or equivalently, the absence of CPW, was missed in \cite{KW}. 

Farther away from the TCP, for $t > t_u^+$, the longest edge of the three-phase triangle (Fig.1) is no longer spanned by  $\alpha$ and $\gamma$ and the necessary condition for wetting by  $\beta$ is no longer met. Instead the longest edge may be spanned by $\alpha$ and $\beta$, in the vicinity of a CEP for $\beta\gamma$ criticality. The wetting phase then becomes the $\gamma$ phase. 

In the three-phase region with $t > t_u^+$ and $s>0$ there is a locus $t\equiv t^{\gamma}_{\rm w} (s)$ of critical (second-order) wetting transitions,
in which $\gamma$ intrudes between $\alpha$ and $\beta$. The wetting phase boundary terminates on the CEP line  at $u^+$ (see Fig.2). The relevant angle for describing (non-)wetting by $\gamma$ is $\hat\gamma$, and it satisfies \eqref{cosbeta} after a cyclic permutation of the phase labels. So we find that CPW takes place in two distinct regimes, separated by a nonwetting gap. Varying $t$, from the TCP outwards, one encounters CPW, nonwetting, and ``reentrant" CPW (Fig.2).  

The nonwetting gap contradicts Cahn's theory and is at variance with the wetting phase diagrams derived for wall-fluid systems with short-range forces \cite{Cahn,NF,PNAS}. This requires a novel explanation. First, notice the three (dotted) lines of symmetry in Fig.2, along each of which two interfacial tensions are equal. These may act as ``neutral lines" with respect to wetting. For example, the line $\sigma_{\alpha\beta}$ = $\sigma_{\alpha\gamma}$ excludes wetting by $\beta$ or $\gamma$, since $\hat\beta=\hat\gamma$. Some exactly calculated points on this line are $(s=0,\, t=1)$, $(s=(1/\sqrt{3})^{3/2}/\sqrt{2}, \,t=1/\sqrt{3})$ and $(s= (1/2)^{3/2}, \,t=1/2)$. The latter is a common point of the neutral line and the CEP line. It is characterized by $\hat \alpha$ = 180$^{\circ} $ and, asymptotically, $\hat \beta = \hat \gamma$ = 90$^{\circ}$. Next, we know that in the wet regimes the asymptotic angle is $\hat \beta = 0$ or $\hat \beta$ = 180$^{\circ}$. If there were no nonwetting gap, the asymptotic value of $\hat \beta$ on the CEP would have to jump discontinuously from $0$ to 180$^{\circ}$ at the neutral point, where, by symmetry, it is 90$^{\circ}$. 

A discontinuity of this caliber in a dihedral angle is possible, and does exist, in the single-density theory with a wall, when the surface field is varied through zero, asymptotically close to the critical temperature $T_c$. This is conspicuous in the wetting phase diagram in \cite{NF}. However, there are no surface fields or other boundary effects that can cause discontinuities in observable dihedral angles, when all three phases are treated on equal footing and only bulk thermodynamic fields are varied within the two-dimensional subspace of three-phase equilibria. Consequently, all angles $0 < \hat \beta <$ 180$^{\circ}$ must be encountered in a continuous manner also when moving along, and infinitesimally close to, the CEP line. This implies the existence of a nonwetting gap.

This general argument is supported in detail by the exact solution of the DFT.  Upon approach of a CEP, the $\beta\gamma$ interface becomes diffuse as $\beta$ and $\gamma$ become one and the same critical phase, named $\beta\gamma$. The angle $\hat\alpha$ approaches 180$^{\circ}$, and $\alpha$ and $\beta\gamma$ each fill a half-space. They are separated by a planar non-critical interface, denoted by $\alpha,\beta\gamma$. For $t_u^- < t < t_u^+$, in the $\beta\gamma$ half-space the diffuse $\beta\gamma$ interface, with vanishing interfacial tension, makes a non-zero contact angle $\hat\beta$ with the $\alpha,\beta\gamma$ interface. The diffuse interface is still bound, or localized, at the non-critical interface in the limit that the CEP is attained. 

From the analytic expressions for the interfacial tensions we calculate, using series expansion to third order in the small distance $\epsilon = 1-s\,t^{-3/2}$ from the CEP line, the asymptotic contact angle $\hat\beta$ along the CEP line. The exact result, displayed in Fig.3, is 
\begin{equation}
\cos\hat\beta = \frac{(2t-1)(1-31t+4t^2)}{((1+t)(1+4t))^{3/2}}, \;\mbox{for}\;t_u^- < t < t_u^+.
 \end{equation}

\begin{figure}[h!]
\centering
\includegraphics[width=0.9\linewidth]{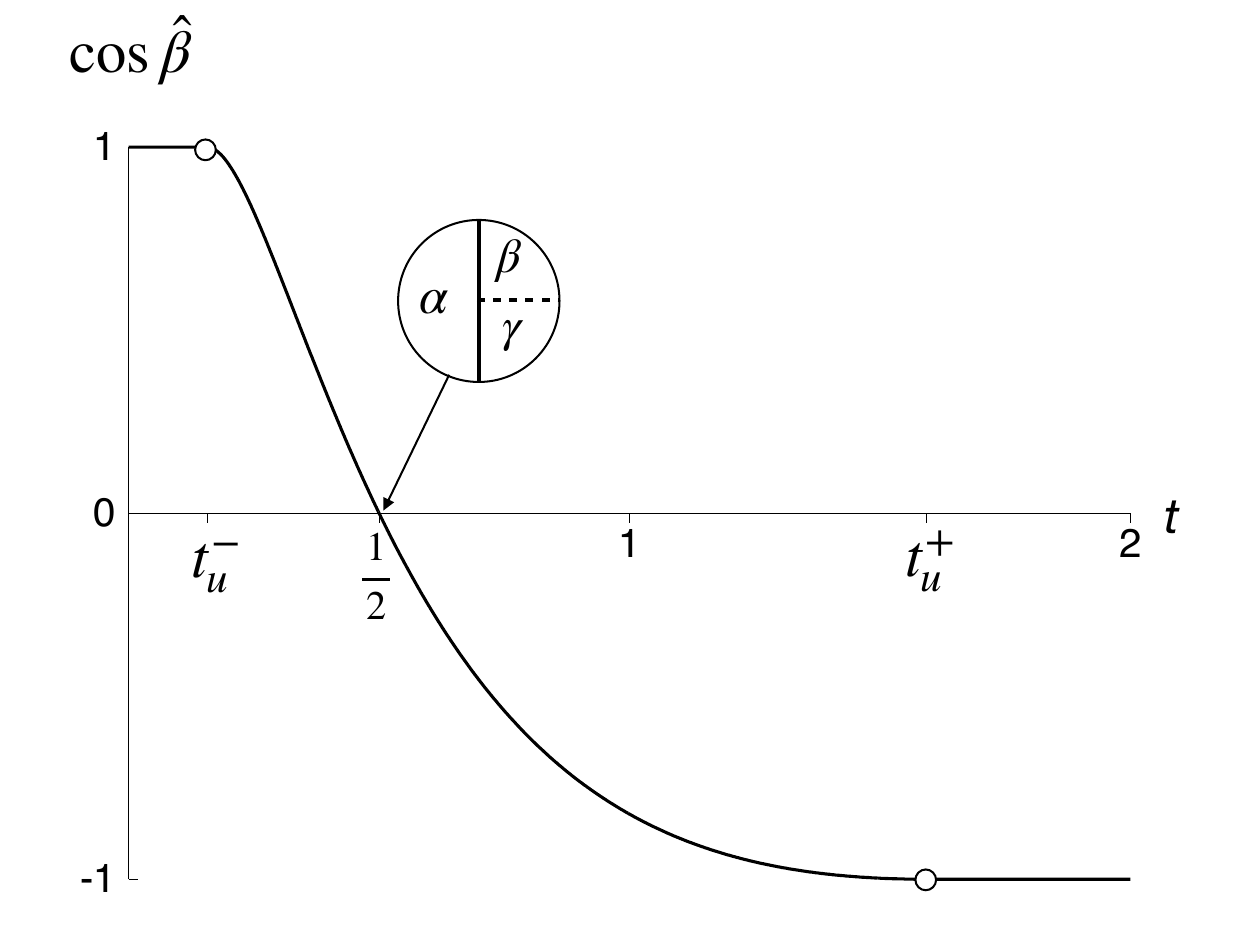}
\caption{Infinitesimally close to the CEP line for $\beta\gamma$ criticality, the cosine of the asymptotic contact angle $\hat \beta$ is a continuous function of the bulk thermodynamic field $t$. Second-order diffuse-interface unbinding occurs at $t = t_u^- \equiv (7 - \sqrt{33})/8$ where $\hat\beta = 0$, $\hat \gamma$ = 180$^{\circ}$ and wetting by $\beta$ terminates at a CEP, and also at $t= t_u^+\equiv (7 + \sqrt{33})/8$ where $\hat\beta$ = 180$^{\circ}$, $\hat \gamma = 0$ and wetting by $\gamma$ terminates at a CEP. The three-phase configuration for $t=1/2$ where a neutral line (see Fig.2) terminates at a CEP is also shown.}
\label{1}
\end{figure}

The values $t_u^-$ and $t_u^+$, which span the nonwetting gap, are found by solving $\cos^2\hat\beta =1$ (which leads to $t(1-7t+4t^2)^2=0$).  When $t \downarrow t_u^- $, $\cos\hat\beta \rightarrow 1$. When $t \uparrow t_u^+$, $\cos\hat\beta \rightarrow -1$ and reentrant CPW is found. 

Starting in the nonwetting gap and moving along, and infinitesimally close to, either one of the two CEP lines, we may encounter an interfacial phase transition in which the diffuse interface unbinds from the planar non-critical interface. These ``diffuse-interface unbinding" transitions, at $u^{\pm }$, are critical and of second order, with 
\begin{equation}
\cos\hat\beta \pm 1 \propto (t-t_u^{\pm })^2, \;\; \mbox{for } \; s \,t^{-3/2} = 1^-,
\end{equation}

Our findings make it necessary to reinterpret the experimental observation of a hitherto unexplained nonwetting state, which continues to exist when either one of the two CEPs is approached in ternary H$_2$O-oil-nonionic amphiphile mixtures at three-phase coexistence \cite{KahlweitBusse}. A drop of the ``middle" phase was observed not to spread but to form a lens, with a contact angle close to 90$^{\circ}$. It was thought that this angle must approach exactly 90$^{\circ}$ close to the CEP based on a scaling argument. Other values, $0 < \hat \beta <$ 180$^{\circ}$, were, unfortunately, not given attention because the authors excluded the very case ($x=y$ in \cite{KahlweitBusse}) which is realized in the nonwetting gap of our exactly solved two-density DFT.   

In conclusion, for three-fluid-phase equilibria in systems with a TCP that are not approximated by two-phase equilibria at a wall, but described by a two-density DFT, the global wetting phase diagram largely contradicts the CPW scenario. A pronounced nonwetting gap is found in an exactly solved DFT paradigm and, we argue, must also generally be present by virtue of thermodynamic continuity of the dihedral angles as a function of bulk thermodynamic field variables, in three-phase equilibria of fluids.

We are grateful to Ben Widom for his insights and interest in this research. We thank Jonas Berx for discussions. We are indebted to the three reviewers for useful remarks.  JOI thanks Okayama University for generous hospitality, the Japan Society for the Promotion of Science for a JSPS Invitational Fellowship for Research in Japan with ID No. S18131 and FWO-Flanders for grant K204422N for a short stay. This work was supported in part by JSPS KAKENHI (Grant Nos. 18KK0151, 20H02696).  

{}
\end{document}